# Asymptotic of 'rigid-body' motions for nonlinear dynamics: physical insight and methodologies[1]

V.N. Pilipchuk

*Wayne State University, Detroit, Michigan  48202*

*Abstract:* The purpose of the present work is to show that an adequate basis for understanding the essentially nonlinear phenomena must also be essentially nonlinear however still simple enough to play the role of a basis.  It is shown that such types of 'elementary' nonlinear models can be revealed by tracking the hidden links between analytical tools of analyses and subgroups of the rigid-body motions or, in other terms, rigid Euclidean transformation.  While the subgroup of rotations is linked with linear and weakly nonlinear vibrations, the translations with reflections can be viewed as a geometrical core of the strongly nonlinear dynamics associated with the so-called vibro-impact behaviors.  It is shown that the corresponding analytical approach develops through non-smooth temporal substitutions generated by the impact models.

## 1.  Introduction

This work is motivated by the intent to introduce a unified physical basis for analyzing vibrations of essentially unharmonic, non-smooth or may be discontinuous time shapes. Transitions to non-smooth limits can make investigations especially difficult due to the fact that the dynamic methods were originally developed within the paradigm of smooth motions based on the classical theory of differential equations. From the physical standpoint, such way is natural for modeling the low-energy motions. Although the impact dynamics has also quite a long pre-history, non-smooth behaviors are often viewed as an exemption rather than a rule. Notice that the classical theory of differential equations usually avoids non-differentiable and discontinuous functions. Presently, however, many theoretical and applied areas are dealing with the high-energy phenomena accompanied by strongly non-linear spatio-temporal behaviors making the classical smooth methods difficult to apply.  For instance, such phenomena are considered in engineering analyses of dynamical systems under constraint conditions, friction-induced vibrations, structural damages due to cracks, liquid sloshing impacts, etc. Similarly to the well-known analogy between mechanical and electrical harmonic oscillators, the so-called Schmitt trigger circuits generate non-smooth signals whose temporal shapes resemble the mechanical vibro-impact processes. In many such cases, it is still possible to adapt different smooth methods of the dynamic analyses through strongly non-linear algebraic manipulations with state vectors or by splitting the phase space into multiple 'smooth domains.' As a result, formulations are often reduced to the discrete mappings in a wide range of the dynamics from periodic to stochastic. It will be shown that a complementary analytical tool can be built on generating models developing essentially unharmonic behaviors as their inherent properties. For instance, the methodology presented in this work, employs elementary impact systems as a physical basis for describing different types of unharmonic processes. This is implemented through the non-smooth time substitutions introduced originally for strongly nonlinear however smooth models [1]. Besides, this tool reveals explicit links between the impact dynamics and the algebra of hyperbolic numbers analogously to the link between harmonic vibrations and conventional (elliptic) complex analyses [2].

---

[1] A keynote lecture at 12th DSTA Conference, December 2-5, 2013 Łódź, Poland

## 2. Physical and mathematical principles

### 2.1. Linear and *elementary nonlinear* phenomena

Although the notion of linearity is quite clear in terms of mathematical formulations, the attempt to directly associate the mathematical definition with physical phenomena faces ambiguities due to the fact that the differential equations of motion for the same model may appear to be either linear or nonlinear if switching between different types of coordinates. For instance, a mechanical system, which is linear in Cartesian coordinates, becomes nonlinear in polar coordinates. Nevertheless, recognizing the unique role of Cartesian coordinates, it was suggested to define a mechanical system as linear if the corresponding differential equations of motion are linear in Cartesian coordinates[2]. The purpose of this subsection is to determine the most elementary dynamic *phenomena* that can already be qualified as essentially nonlinear. For that reason, it is convenient to consider the linear situation first, for instance, on a typical mass-spring vibrating model; see figure 1 (a), where all the springs are assumed to be linearly elastic. The corresponding position vector is represented in the following complex form

$$\overline{q} = A_1 \exp(i\omega_1 t + \varphi_1) + A_2 \exp(i\omega_2 t + \varphi_2) \qquad (1)$$

where $A_k$ are constant complex vectors, $\omega_k$ and $\varphi_k$ ($k = 1,2$) are the modal frequencies and initial phases.

The fragment (b) in figure 1 shows another mechanical system whose dynamics is described by the same expression (1), if the system' position is identified by the two points fixed at the edges of the discs. Therefore, the one-dimensional dynamics generated by linearly elastic restoring forces can be represented by the rigid-body rotations. In other words, the linearly elastic forces are effectively eliminated by increasing the dimension of space. To some extent, such simple observation provides an explanation why the sine and cosine waves possess so convenient mathematical properties. Namely these harmonic functions represent the subgroup of elementary rigid-body motions, such as rotation. Furthermore, the link between the two models, (a) and (b), in figure (1), enables one to associate the linearity with the subgroup of rotations.

---

[2] This definition was suggested by V. Zhuravlev (private communication, Moscow, 1989.)

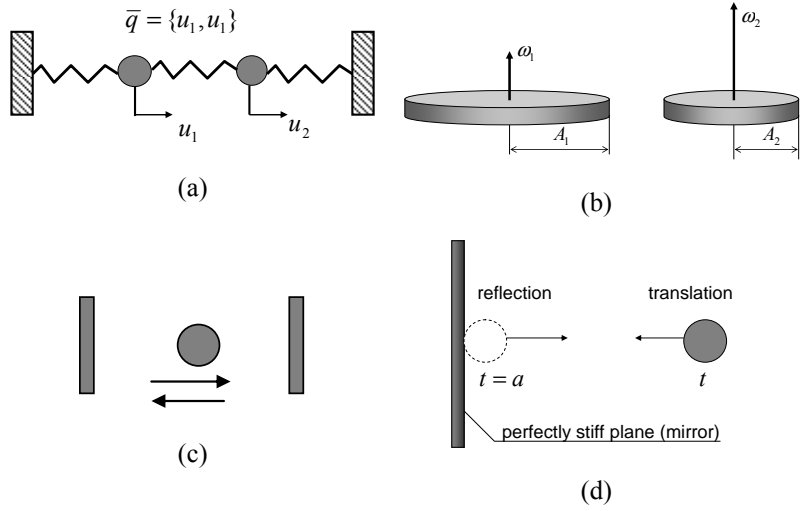

**Figure 1.** Vibration of the linearly elastic system (a) is represented by the rigid-body rotations (b); the impact oscillator (c) is associated with the rigid-body translation and reflection (d).

| | **Mechanical model** | **Basic function** | **Substitution** |
|---|---|---|---|
| 1 | $t=a$, $v=1$, $s(t)$ | $s = \lvert t-a \rvert$ | $t = a + s\dot{s}$ <br> $\dot{s}^2 = 1$ |
| 2 | length $1 \mid 1$, $v=1$, $\tau(t)$ | $\tau = \dfrac{2}{\pi}\arcsin\sin\left(\dfrac{\pi t}{2}\right)$ | $t = 1 + (\tau-1)\dot{\tau}$ <br> $\dot{\tau}^2 = 1$ |
| 3 | $v=1$, $x=0$, $s(t;d)$, $x=1$ | $s(t;d) = \dfrac{1}{2}(d + \lvert t \rvert - \lvert t-d \rvert)$ | $t = \sum_{i=0}^{\infty}(t_i + s_i)\dot{s}_i$ <br> $\dot{s}_i^2 = 1$ |

**Figure 2.** Different versions of non-smooth temporal substitutions generated by the impact models.

Apparently, the rotations do not cover all the rigid-body motions – rigid Euclidean transformations. These include also translations and reflections. The translation itself seems to have a little physical content. However, combinations of translation and reflection appear to be more interesting as shown in figure 1 (c) and (d) through the corresponding mechanical representations. In particular, the fragment (c) shows a typical impact oscillator, whereas the fragment (d) illustrates a single reflection case, which can be viewed as the most elementary nonlinear phenomenon.

**2.2. Impact models and non-smooth temporal substitutions**

Based on the class of linear generating models, whose typical representative is shown in figure 1 (a), the quasi-harmonic methods for nonlinear vibration theory are well developed and widely used. In contrast, using the models associated with the translation-reflection, as those shown in figure 1 (c-d), is less common. Despite of being geometrically simple, such models are not described within the classical theory of differential equations due to the presence of discontinuities in the dynamic states. This fact essentially complicates any direct use of the impact models as a basis for building asymptotic or iterative procedures. It is shown, however, that appropriate preliminary adaptations of the differential equations of motion can be conducted through the non-smooth time substitutions listed in the third column of figure 2. In particular, the first row explains how such type of substitutions is introduced. Namely, the basic function $s = |t - a|$ is the model' coordinate, which can be interpreted as an *eigentime* of the system provided that nothing else happens except the translation and reflection under consideration. The goal is to introduce a new temporal argument, say $s$, in order to obtain the differential equation of motion for the model, which is shown on the left of first row in figure 2. Apparently, the substitution $t \rightarrow s$ cannot work directly since no inversion $t = t(s)$ does exist on the entire time domain. Nevertheless, the following generalization holds $t = t(s, \dot{s})$; see the first row on its right in figure 2 for details. In particular, this inversion appears to have the specific algebraic structure of hyperbolic numbers with the basis $\{1, \dot{s}\}$. In contrast to the conventional elliptic complex numbers, the *unipotent* $\dot{s}$, if squared, gives the positive sign namely $\dot{s}^2 = 1$. Interestingly enough, the hyperbolic numbers have been known for about one and a half century as abstract algebraic elements with no relation to the nonlinear dynamics or non-smooth functions. Moreover, such numbers have been re-introduced multiple times under different names; see [3, 4] for an explanation of this fact and further details. In the present case, the hyperbolic structures are generated by the non-smooth temporal arguments of impact systems, while the unipotent $\dot{s}$ possesses a clear physical interpretation as the normalized velocity of impact system. Different applications to the vibration problems are based on the periodic version [1, 2] illustrated by the second row of figure 2. In particular, the equation on the right of the row means that, *during one period, the time argument is expressed through the dynamic states of impact oscillator in the form of hyperbolic number with the unipotent* $\dot{\tau}$. This statement therefore applies to any periodic process as was originally shown in [1]. Figure 3 provides a geometrical interpretation of the 'oscillating time' $\tau$ for the particular case of periodic process described by the even function $x(t)$ with respect to the quarter of period, $t = a$.

Finally, in the third row of figure 2, the original time argument is structurised to match the one-dimensional dynamics of rigid-body chain of identical particles. The continuous 'global' time is associated with the propagation of linear momentum, whereas a sequence of non-smooth 'local' time arguments describes the behavior of individual physical particles. Such an idea helps to incorporate the temporal symmetries of the dynamics into the differential equations of motion in many other cases of regular or irregular sequences of internal impacts or external pulses. Since the local times are bounded, a wider range of analytical tools becomes applicable.

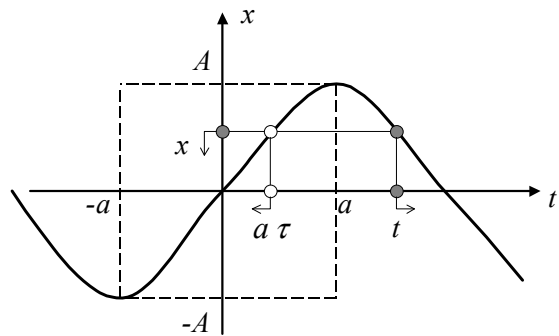

**Figure 3.** The 'oscillating time' τ changes its direction whenever a system makes a *U* – turn, while the original physical time *t* runs to infinity.

### 2.3. Different asymptotic approaches to the vibration theory

Figure 4 provides further illustration for logical links between two alternative approaches to the vibration problems. Interestingly enough, the illustration is possible within the same one-degree-of-freedom model, which is shown at the first row of figure 4. Note that the oscillators with power-form characteristics were considered for quite a long time. For instance, Lyapunov obtained such oscillators while investigating degenerated cases of the dynamic stability problems. Besides, he introduced a couple of special functions, *cs* and *sn*, in order to invert the corresponding quadratures [5]. This relatively simple model nevertheless depicts the gradual transition from linear to strongly nonlinear dynamics as the exponent *n* runs from unity to infinity. Notably, all the temporal mode shapes of the oscillator are described by the special functions *cs* and *sn*, except the two boundaries of the inter-val $1 \leq n < \infty$. Both boundaries represent simple asymptotic limits described within the class of elementary functions.

| | **Harmonic limit** | **Impact limit** |
|---|---|---|
| 1 | 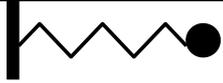<br>$\ddot{x} + x^{2n-1} = 0, \quad n = 1$<br>*Harmonic oscillator* | 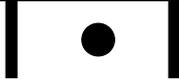<br>$\ddot{x} + x^{2n-1} = 0, \quad n \to \infty$<br>*Impact oscillator* |
| 2 | 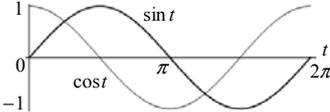<br>*Sine & cosine waves*<br>*Subgroup of rotations* | 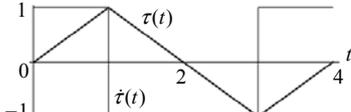<br>*Triangular sine wave*<br>*Translation-reflection* |
| 3 | $z = x + i\dot{x}$<br>$i^2 = -1$<br>*Elliptic complex numbers* | $x = X(\tau) + Y(\tau)\dot{\tau}$<br>$\dot{\tau}^2 = 1$<br>*Hyperbolic complex numbers* |
| 4 | 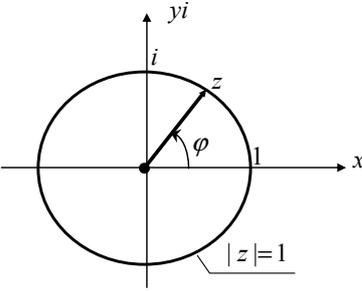<br>*Elliptic complex plane* | 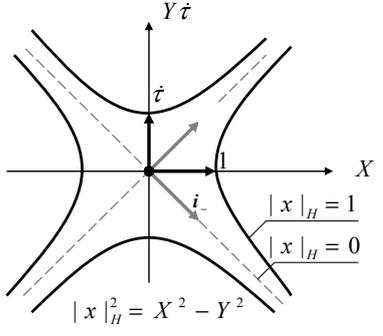<br>*Hyperbolic complex plane* |
| 5 | $\sum_{k=0}^{\infty} A_k \cos kt + B_k \sin kt$<br>*Fourier series* | $\sum_{k=0}^{\infty} \frac{1}{k!} X^{(k)}(0) \tau^k + \frac{1}{k!} Y^{(k)}(0) \tau^k \dot{\tau}$<br>*Saw-tooth power series* |

**Figure 4.** Two alternative approaches to the vibration theory based on the harmonic and impact limits.

Consider first the limit of harmonic oscillator ($n = 1$), generating the *sine* and *cosine* waves; see the left of the second row. The widely known convenience of using this couple of functions can be explained by their links to the elementary rigid-body motions namely the subgroup of rotations. The algebra of conventional (elliptic) complex numbers with the corresponding complex plane representation can be viewed as a next hierarchic step here due to the well-known Euler formula. Finally, taking the linear combination of harmonic waves with different frequencies and keeping in mind the idea of parameter variations leads the area of harmonic and quasi harmonic analyses of vibrating systems. Such tools therefore represent the dynamic processes as a combination of the elementary rigid-body rotations with different angular speeds.

Let us consider now the limit $n \rightarrow \infty$, when the restoring force vanishes inside the interval $-1 < x < 1$ but becomes infinitely large as the system reaches the potential barriers at $x = \pm 1$. The physical meaning of this limit is introduced at the top of the right column in figure 4. Despite of the strong (impact) nonlinearity, the limiting oscillator is also described by quite simple elementary functions such as the triangular sine and rectangular cosine, say $\tau$ and $\hat{\tau}$. These two non-smooth functions are associated with another subgroup of the rigid-body motions namely translation and reflection. Therefore, analogously to the case $n = 1$, the upper limit $n = \infty$ can play the same fundamental role by generating a hierarchy of tools as listed in the right column of figure 4.

## 3. Further mathematical properties and examples

This section describes the basic mathematical properties of the non-smooth temporal substitutions introduced in figure 2. These properties are used then for derivations in different illustrating examples.

### 3.1. Mathematical properties

Consider first, the single reflection case; see the first row of figure 2. Algebraic, differential, and integral properties are as follows:

- Isomorphism with 2×2 symmetric matrixes

$$t^2 = (a + s\dot{s})^2 = a^2 + s^2 + 2as\dot{s}$$

$$\hat{t}^2 = \begin{pmatrix} a & s \\ s & a \end{pmatrix}^2 = \begin{pmatrix} a^2 + s^2 & 2as \\ 2as & a^2 + s^2 \end{pmatrix}$$

- Functional linearity holds for any function *x(t)*:

$$x(t) = x(a + s\dot{s}) = X(s) + Y(s)\dot{s} \qquad (2)$$

$$X(s) = \frac{1}{2}[x(a+s) + x(a-s)]$$

$$Y(s) = \frac{1}{2}[x(a+s) - x(a-s)]$$

For instance, $\exp(t) = \exp(a)(\cosh s + \dot{s} \sinh s)$

- Division is conditioned to exclude the possibility of zero denominators:

$$t^{-1} = \frac{(a - s\dot{s})}{(a + s\dot{s})(a - s\dot{s})} = \frac{a - s\dot{s}}{a^2 - s^2} = \frac{a}{a^2 - s^2} - \frac{s}{a^2 - s^2}\dot{s}$$

$$(s \neq |a|)$$

- Sequential differentiation remains in the algebra of hyperbolic numbers under the smoothness conditions at $s = 0$:

$$\frac{d}{dt}[X(s) + Y(s)\dot{s}] = Y'(s) + X'(s)\dot{s} \qquad (3)$$
$$\text{if} \quad Y(0) = 0$$

$$\frac{d^2}{dt^2}[X(s) + Y(s)\dot{s}] = X''(s) + Y''(s)\dot{s}$$
$$\text{if} \quad Y(0) = 0, \quad X'(0) = 0$$

- Integration remains in the algebra of hyperbolic numbers:

$$\int [X(s(t)) + Y(s(t))\dot{s}(t)] dt$$
$$= \left[\int_0^s Y(z) dz + C\right] + \left[\int_0^s X(z) dz\right]\dot{s}$$

### 3.2. Sample solution procedure for the *s*-case

Let us consider the following initial value problem

$$\dot{x} + \lambda x = 2p\delta(t-a) \equiv p\ddot{s}$$
$$x(0) = 0 \qquad (4)$$
$$s = |t-a|$$

Introducing the new temporal argument, $t \to s$, generates the following substitution for the unknown function: $x(t) = X(s) + Y(s)\dot{s}$. As a result, the differential equation takes the form

$$\underbrace{Y' + \lambda X}_{\text{regular}} + \underbrace{(X' + \lambda Y)\dot{s}}_{\text{step-wise discont.}} + \underbrace{(Y-p)\ddot{s}}_{\text{singular term}} = 0 \qquad (5)$$

Equating separately the terms of different level of singularity in (5) to zero leads to the following boundary value problem with *no singular terms*

$$\begin{cases} Y' + \lambda X = 0 \\ X' + \lambda Y = 0 \\ p - Y(0) = 0 \end{cases} \qquad (6)$$

The initial condition in (4) yields

$$X(a) - Y(a) = 0 \qquad (7)$$

The boundary value problem (6)-(7) is easy to solve in few steps. Then the corresponding solution of the original initial value problem (4) is obtained in the closed-form $x(t) = p\exp(-\lambda s)(1+\dot{s})$; see figure 5 for illustration.

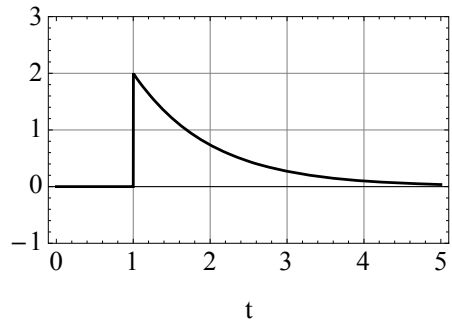

**Figure 5.** Solution of problem (4) under the following parameters: *P = 1.0; a = 1.0; λ = 1.0*.

### 3.3. Properties and sample solution for the τ-case

Algebraic and differential properties of the periodic τ-version, which is shown in the second row of figure 2, are similar to those listed in section 3.1. Note that, although the analytical definitions for the basic functions $\tau$ and $\dot{\tau}$ look more complicated in this case, there is no need for memorizing them. All what is necessary for solving problems is the following properties

$$\dot{\tau} = e, \quad e^2 = 1, \quad \dot{e} \neq 0 \Leftrightarrow \tau = \pm 1 \tag{8}$$

The third relationship in (8) means that, whenever τ reaches its amplitude values, a δ-spike occurs from the following series

$$\dot{e}(t) = 2 \sum_{k=-\infty}^{\infty} \left[ \delta(t+1-4k) - \delta(t-1-4k) \right]$$

In other words, the present situation is quite similar to analytical manipulations with the conventional trigonometric functions using only the function properties with no involvement of their definitions. In order to illustrate the manipulations, let us consider the overdamped oscillator under the rectangular cosine loading

$$\dot{x} + \lambda x = p e(t) \tag{9}$$

The unknown periodic solution is represented in the form [1, 2]

$$x = X(\tau) + Y(\tau)\dot{\tau} \tag{10}$$

Substituting (10) in (9) and taking into account properties (8) gives

$$\underbrace{Y' + \lambda X}_{\text{regular}} + \underbrace{(X' + \lambda Y - p)e}_{\text{step-wise discont.}} + \underbrace{Y\dot{e}}_{\text{singular}} = 0 \tag{11}$$

Equating separately to zero the terms of different level of singularity gives the autonomous boundary value problem with no discontinuities

$$\begin{cases} X' + \lambda Y = p \\ Y' + \lambda X = 0 \\ Y(\pm 1) = 0 \end{cases} \tag{12}$$

Substituting solution of the boundary value problem (12) in (10) gives finally the periodic closed form solution of the original equation (9); see figure 6 for illustration:

$$x = \frac{p}{\lambda}\left\{\frac{\sinh(\lambda\tau)}{\cosh\lambda} + \left[1 - \frac{\cosh(\lambda\tau)}{\cosh\lambda}\right]e\right\} \quad (13)$$

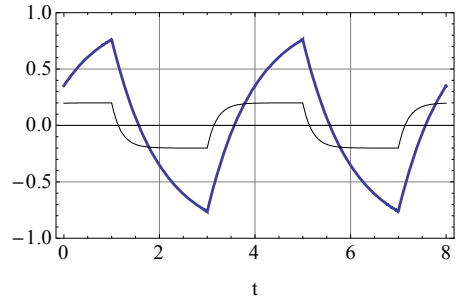

**Figure 6.** Solution (13) for two different magnitudes of the parameter: $\lambda = 1.0$ – solid line, larger amplitude; $\lambda = 5.0$ – thin line, smaller amplitude.

Obviously, equation (9) can be also solved by means of either Fourier series or Laplace transforms, or directly by matching different pieces of solution under periodicity conditions. However, using the Fourier series, for instance, requires a very large number of terms near the non-smoothness points as shown in figure 6.

### 4. Conclusions

This paper outlines the basic principles of non-smooth temporal substitutions and gives exactly solvable illustrating examples. Note that the transition to non-smooth temporal argument should be viewed as a preliminary stage of analyses. Such stage of transformation incorporates the specifics of external loading or/and inherent properties of a physical system into the new equations. As a result, a much wider range of analytical and numerical methods becomes possible to apply. This is due to the fact that the new temporal arguments vary within narrower domains and automatically capture the major temporal symmetries of the dynamics, such periodicity and reflections. The corresponding analytical algorithms and solutions for different strongly nonlinear oscillators can be found in the reference [2]. The typical form of such solutions is power series with respect to the triangular sine $\tau$; see row 5 in figure 4 for explanation. Note that direct power series expansions with respect to the original time $t$ usually make little sense for vibration problems due to the loss of periodicity. The amplitude and frequency modulated motions can be analyzed by adding a slow time argument to representation (10) and then using the idea of two variables or multiple scale expansions. Let us men-tion also a new area of applications, which is being developed due to an interesting observation that the temporal mode shapes of the phase variable, describing the 1:1 resonance energy exchange between weakly nonlinear oscillators, resembles the dynamic states of impact oscillator [6]. In particular, it was found that such "impacts" take place when the entire energy is involve into the exchange process.

Finally, let us summarize the revealed links between specific cases of the Euclidean rigid transformations, the induced algebraic structures, and linear and nonlinear dynamics. The class of rigid transformations $T$ of an arbitrary vector $r$ is described by the expression $T(r) = Ar + b$, where $A$ and $b$ are the orthogonal matrix and a constant translation vector, respectively. Then the above mentioned logical links are illustrated by the table, which can be viewed as an extension of figure 4,

$$\det(A) = \begin{cases} +1 & \Leftrightarrow \text{Rotation} \quad \Leftrightarrow i^2 = -1 \quad \text{Elliptic numbers} \quad \Leftrightarrow \text{Linear dynamics} \\ 0 & \Leftrightarrow \text{Singular case} \quad \Leftrightarrow i^2 = 0 \quad \text{Parabolic numbers} \quad \Leftrightarrow \quad ? \\ -1 & \Leftrightarrow \text{Reflection} \quad \Leftrightarrow i^2 = +1 \quad \text{Hyperbolic numbers} \Leftrightarrow \text{Nonlinear dynamics} \end{cases}$$

where the singular case generates an open question.

**References**

1. Pilipchuk V.N.: Transformation of oscillating systems by means of a pair of nonsmooth periodic functions. *Dokl. Akad. Nauk Ukrain. SSR,* Ser. A (4), 1988, 37–40.
2. Pilipchuk V.N.: *Nonlinear dynamics: between linear and impact limits.* Berlin, Springer, 2010.
3. Kisil V.V.: Induced representations and hypercomplex numbers. *Advances in Applied Clifford Algebras,* 23(2), 2013, 417-440.
4. http://en.wikipedia.org/wiki/Split-complex_number
5. Lyapunov A.M.: Investigation of a singular case of the problem of stability of motion. *Math. Sbornik*, 17, 1893, 252-333.
6. Manevitch L.I.: New approach to beating phenomenon in coupled nonlinear oscillatory chains. *Arch. Appl. Mech.,* **77**, 2007, 301-312.